\documentclass[a4paper,12pt,notitlepage]{iopart}

\usepackage{iopams}  
\usepackage{hyperref}
\usepackage{enumerate}
\usepackage{color}
\usepackage{graphics}
\usepackage{epsfig}
\usepackage{lineno}
\graphicspath{{./pics/}}

\newif\ifcomment
\usepackage[small,bf]{caption}
\newcommand{\ZDC}          {\rm{ZDC}}
\newcommand{\ZDCs}         {\rm{ZDCs}}
\newcommand{\SPD}          {\rm{SPD}}

\newcommand{\TPC}          {\rm{TPC}}
\newcommand{\VZERO}        {\rm{VZERO}}
\newcommand{\VZEROA}       {\rm{VZERO-A}}
\newcommand{\VZEROC}       {\rm{VZERO-C}}

\newcommand{\pp}           {pp}
\newcommand{\ppbar}        {\mbox{$\mathrm {p\overline{p}}$}}
\newcommand{\PbPb}         {\mbox{Pb--Pb}}
\newcommand{\AuAu}         {\mbox{Au--Au}}

\newcommand{\dNdeta}       {\mathrm{d}N_\mathrm{ch}/\mathrm{d}\eta}
\newcommand{\dEdeta}       {\mathrm{d}E_\mathrm{T}/\mathrm{d}\eta}

\newcommand{\snn}          {\ensuremath{\sqrt{s_{\rm NN}}}}
\newcommand{\snnbf}        {\ensuremath{\mathbf{{\sqrt{s_{\mathbf NN}}}}}}

\newcommand{\Npart}        {\ensuremath{N_\mathrm{part}}}

\newcommand{\Ncoll}        {\ensuremath{N_\mathrm{coll}}}

\newcommand{\abs}[1]       {\ensuremath{\left|#1\right|}}

\newcommand{\Fig}[1]       {Fig.~\ref{#1}}

\begin{document}
\title[Charged-particle multiplicity and transverse energy with ALICE]
      {Charged-particle multiplicity and transverse energy 
       in Pb--Pb collisions at $\snnbf\mathbf{=2.76}$~TeV with ALICE}

\author{Constantin Loizides for the ALICE collaboration}

\address{Lawrence Berkeley National Laboratory, 1 Cyclotron Rd, Berkeley, CA 97420, USA}
\ead{cloizides@lbl.gov}

\begin{abstract}
The measurements of charged-particle multiplicity and transverse energy at mid-rapidity 
in \PbPb\ collisions at $\snn = 2.76$~TeV are reported as a function of centrality. 
The fraction of the inelastic cross section recorded by the ALICE detector 
is estimated using a Glauber model.
The results scaled by the number of participating nucleons are compared with \pp\ collisions 
at the same collision energy, to similar results obtained at significantly lower energies, 
and with models based on different mechanisms for particle production in nuclear collisions. 
\end{abstract}


Strongly interacting matter at extreme energy and matter density over a large volume is studied 
by colliding large nuclei at ultra-relativistic energies.
The multiplicity of charged particles and their transverse energy produced in the central 
rapidity region are fundamental observables to characterize properties, such as the
the initial gluon and energy density, of the matter created in these collisions.
We present the first measurements of the multiplicity density of charged primary particles 
$\dNdeta$~(published in \cite{Aamodt:2010pb,Aamodt:2010cz}) and the transverse energy 
$\dEdeta$ (preliminary) at mid-rapidity in \PbPb\ collisions at $\snn = 2.76$~TeV.
The centrality dependence of these measurements, over nine centrality classes covering the most central 80\% 
of the hadronic cross section, is characterized by the number of participants $\Npart$ determined with a 
Glauber model of \PbPb\ collisions.

The data for these measurements were collected with the ALICE detector in the first \PbPb\ run at the 
Large Hadron Collider~(LHC), November 2010.
The main detectors utilized in the analyses presented here are the Inner Tracking System~(ITS),
in particular its first two layers, the Silicon Pixel Detector (\SPD), and the Time Projection Chamber~(\TPC), 
the main detector for charged particle reconstruction and identification~(via $dE/dx$) at $-0.9<\eta<0.9$.
The \SPD\ consists of two cylindrical layers of hybrid silicon pixel assemblies covering
$|\eta|<2.0$ and $|\eta|<1.4$ for the inner and outer layers, respectively.
The \VZERO\ counters are two arrays of 32 scintillator tiles, covering $2.8<\eta<5.1$~(\VZEROA) 
and $-3.7<\eta<-1.7$~(\VZEROC), and provide both amplitude and time information.
The trigger was configured for high efficiency to accept hadronic events and was successively tightened 
throughout the run period, relying on a combination of the following conditions:
i) two pixel hits in the outer layer of the \SPD, 
ii) a signal in \VZEROA, 
iii) a signal in \VZEROC.

Electromagnetically induced interactions are reduced by requiring an energy deposition above $500$~GeV 
in each of the neutron Zero Degree Calorimeters~(\ZDCs) positioned at $\pm$~114~m from the interaction point. 
Beam background events are removed using the \VZERO\ and \ZDC\ timing information.
The combined trigger and selection efficiency estimated from a variety of Monte Carlo~(MC) studies ranges 
between 97\% and 99\% with a purity of 100\% for the centrality range considered here.

The anchor point for the determination of the $0$--$80$\% most central events is obtained by fitting the minimum bias 
distributions of various detector responses (\VZERO\ amplitudes, \SPD\ outer-layer hits,
or \TPC\ tracks) by a model of particle production based on a Glauber description of 
nuclear collisions.
The model assumes that the number of particle-producing sources is given by 
$f\times\Npart + (1-f)\times\Ncoll$, where $\Npart$ is the number of participating nucleons, 
$\Ncoll$ is the number of binary nucleon--nucleon collisions and $f_i$ and $\alpha_i$ quantify their relative 
contributions.
The number of particles produced by each source is distributed according to a negative binomial distribution.
The nuclear density for $^{208}$Pb is modeled by a 2-parameter Fermi distribution with a radius of $6.62$~fm 
and a skin depth of $0.546$~fm.
A hard-sphere exclusion distance of $0.4$~fm between nucleons is employed.
Nucleons from each nucleus interact if their transverse distance is less than that given by the
inelastic nucleon--nucleon cross section, extrapolated as $64\pm 5$~mb at $\sqrt{s} = 2.76$~TeV, 
consistent with the preliminary measurement $62.1\pm1.6\pm4.3$ mb~\cite{martin}.

The charged-particle multiplicity $\dNdeta$ at $\abs{\eta}<0.5$ is measured counting tracklets, 
defined as a pair of hits, one in each SPD layer, corrected by applying 
a factor $\alpha\times(1-\beta)$ in bins of pseudo-rapidity and $z$-position of the primary vertex.
The factor $\alpha$ is estimated from MC simulations to be about $1.8$, and primarily accounts for the 
acceptance and efficiency,  
while $\beta$ accounts for the fraction of background tracklets from uncorrelated hits, 
and is found to be between $1$\%~(most peripheral events) and $14$\%~(most central events).

\begin{figure}[tbh!f]
\begin{minipage}[t]{0.45\linewidth}
\centering
\includegraphics[width=\textwidth]{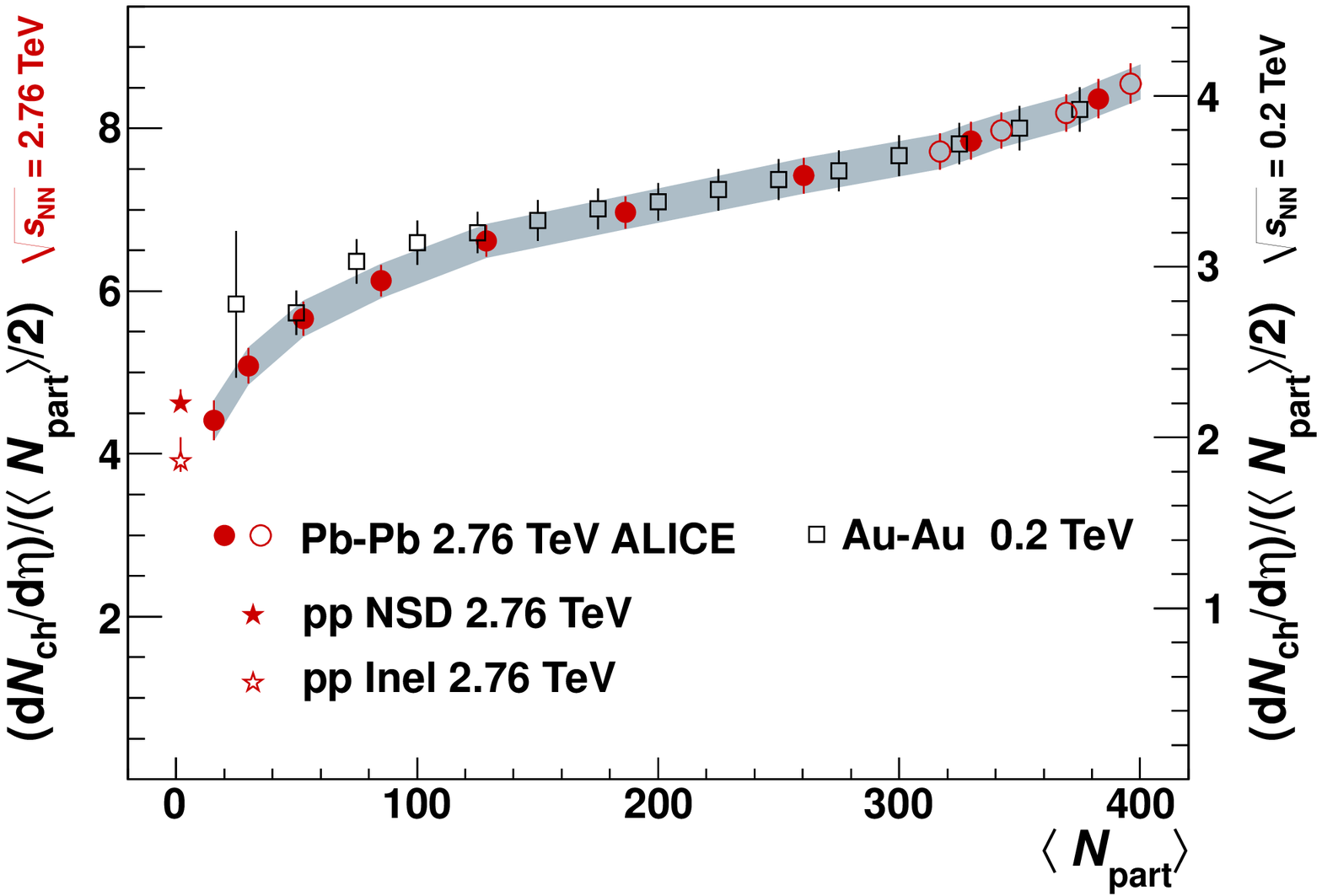}
\caption{
Centrality dependence of the charged-particle pseudo-rapidity density per participant pair for \PbPb\ 
collisions at $\snn=2.76$~TeV and \AuAu\ collisions at $\snn=0.2$~TeV~(RHIC average).
The scale for the lower-energy data is shown on the right-hand side and differs from the scale for the 
higher-energy data on the left-hand side by a factor of 2.1.
The figure is from \cite{Aamodt:2010cz}, where the references to all data points can be found.}
\label{fig1}
\end{minipage}
\hspace{0.5cm}
\begin{minipage}[t]{0.45\linewidth}
\centering
\includegraphics[width=\textwidth]{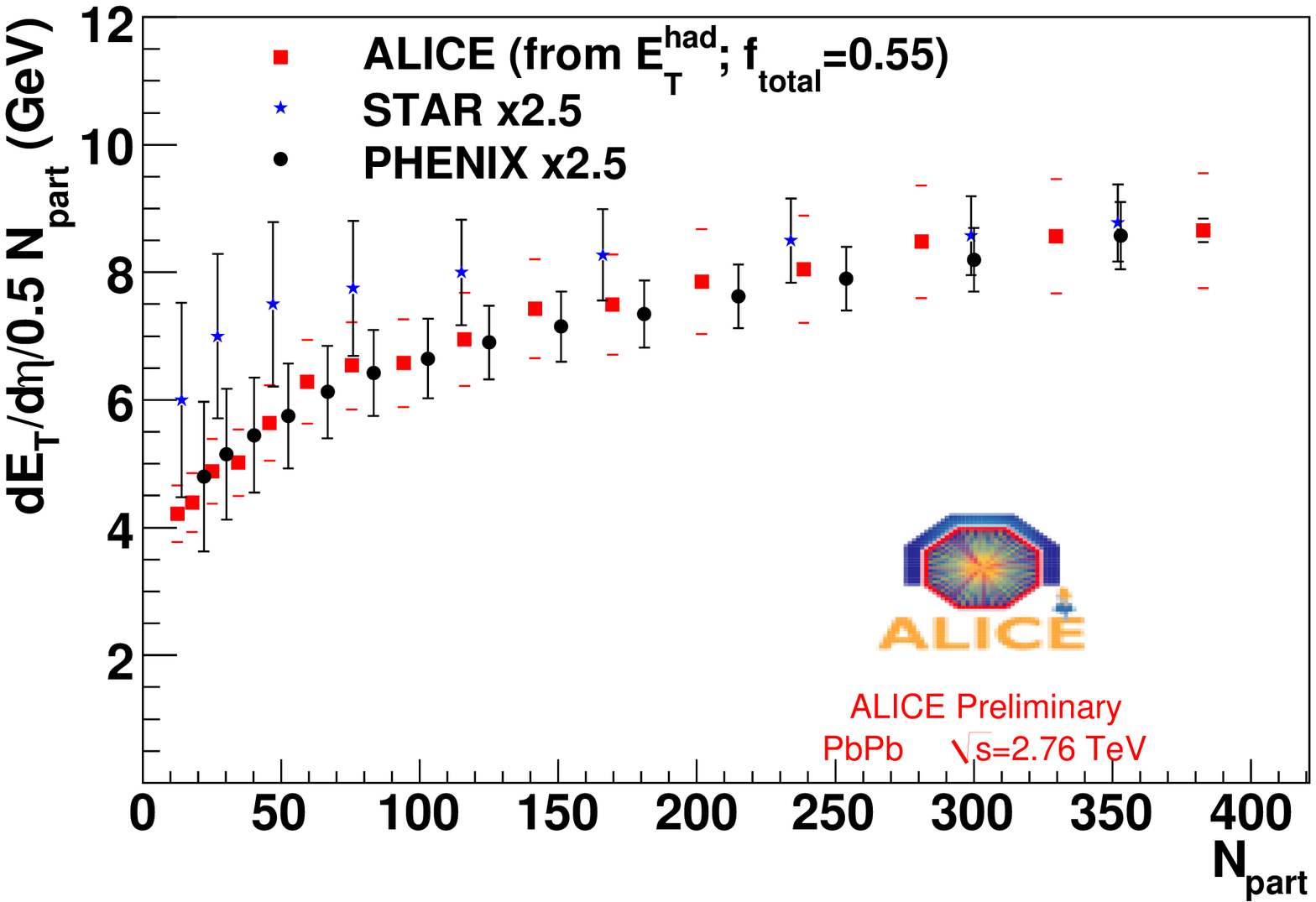}
\caption{
Centrality dependence of the transverse energy per participant pair for \PbPb\ 
collisions at $\snn=2.76$~TeV and \AuAu\ collisions at $\snn=0.2$~TeV~(\cite{Adler:2004zn,Adams:2004cb}).
The lower-energy data are scaled by a factor of 2.5.
}
\label{fig2}
\end{minipage}
\end{figure}

\begin{figure}[tbh!f]
\begin{minipage}[t]{0.47\linewidth}
\centering
\includegraphics[width=\textwidth]{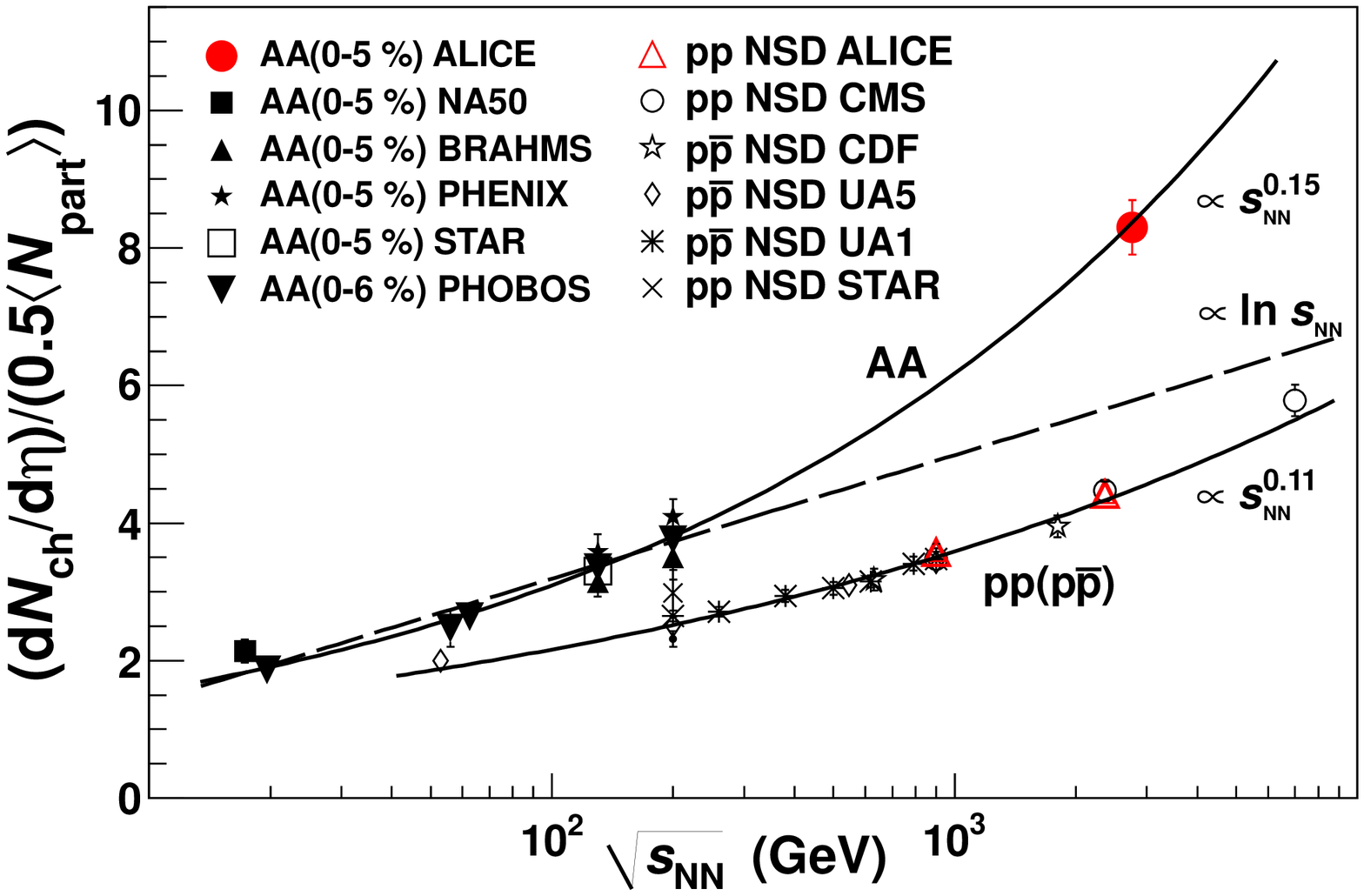}
\caption{
Charged-particle pseudo-rapidity density per participant pair for central nucleus--nucleus and non-single 
diffractive \pp~(\ppbar) collisions as a function of $\snn$. 
The figure is from \cite{Aamodt:2010pb}, where the references to all data points can be found.}
\label{fig3}
\end{minipage}
\hspace{0.5cm}
\begin{minipage}[t]{0.47\linewidth}
\centering
\includegraphics[width=\textwidth]{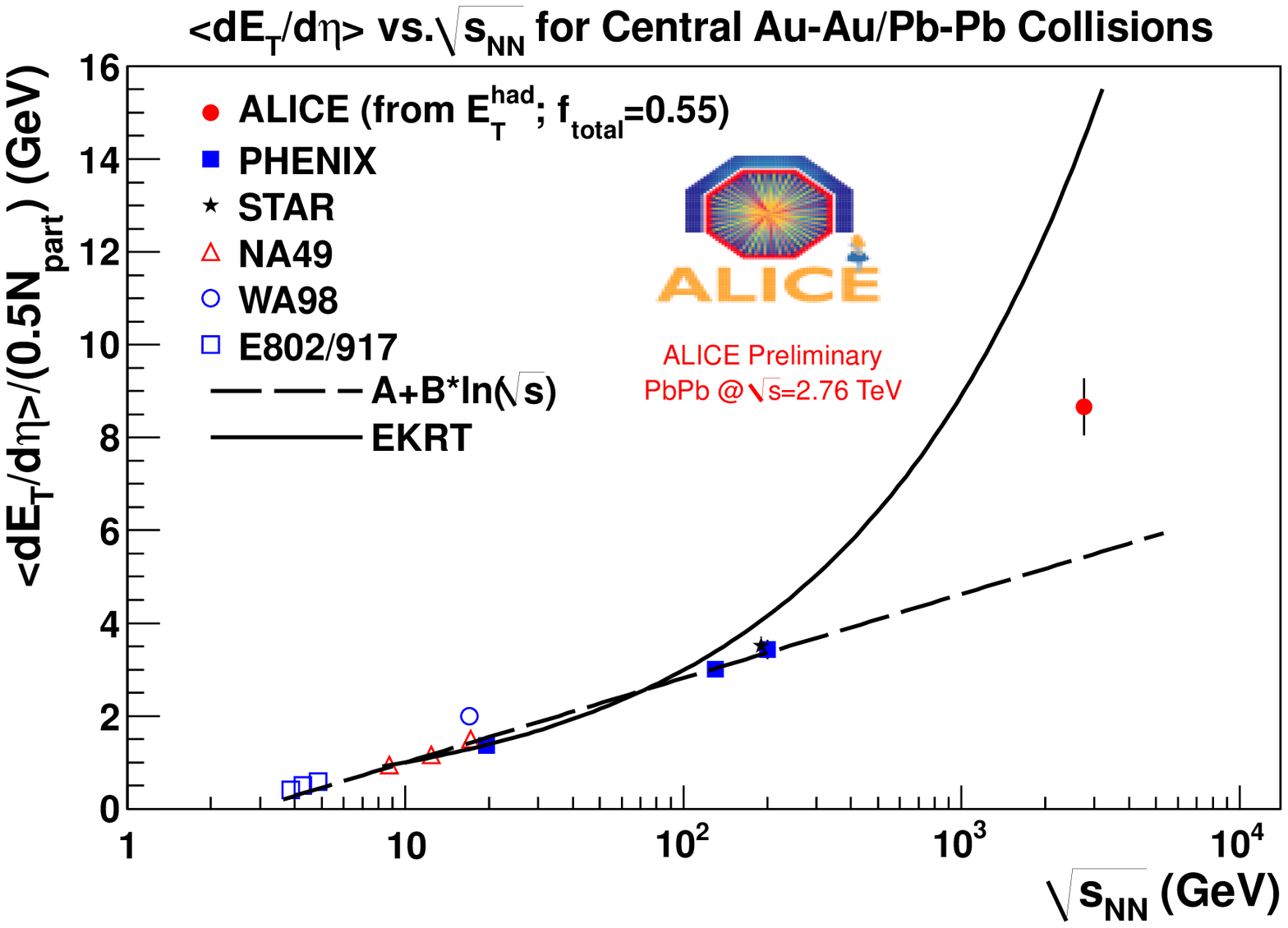}
\caption{
Transverse energy per participant pair for central nucleus--nucleus collisions as a function of $\snn$. 
The data are compared to predictions from EKRT~\cite{Eskola:2001bf} and an extrapolation from lower energy data.}
\label{fig4}
\end{minipage}
\end{figure}

Transverse energy $\dEdeta$ at $\abs{\eta}<0.7$ is obtained by measuring the hadronic energy 
$E^{\rm had}_{\rm T}$ using the barrel tracking detectors corrected by the fraction of particles 
$f_{\rm total}$ not accessible by the tracking detectors. 
MC generators determine that about 91\% of the electro-magnetic contribution to $E_{\rm T}$ arise from 
neutral pions whose transverse energy is approximated by $0.5 E^{\pi^{\pm}}_{\rm T}$.
MC generators typically underestimate the yield of strange hadrons, even in \pp\ collisions.
Therefore, their contribution is derived from L\'{e}vy fits the to \pp\ data at $\sqrt{s}=0.9$~TeV.
The effects of baryon enhancement and strangeness enhancement are estimated by investigating the
data at $\snn=0.2$~TeV.
The total correction~(for $\Lambda$, $k^{0}_{\rm s}$, $n$, $\pi^{0}$, $\eta$ and $\omega$) 
amounts to $f_{\rm total}=0.55\pm0.02$. Other correction factors are typically much 
smaller,and include corrections for finite acceptance at low transverse momentum and tracking efficiency, 
as well as corrections for particles which can not be identified via $dE/dx$ and for contributions of 
mis-identified primary particles.

\begin{figure}[tbh!f]
\begin{minipage}[t]{0.48\linewidth}
\centering
\includegraphics[width=\textwidth]{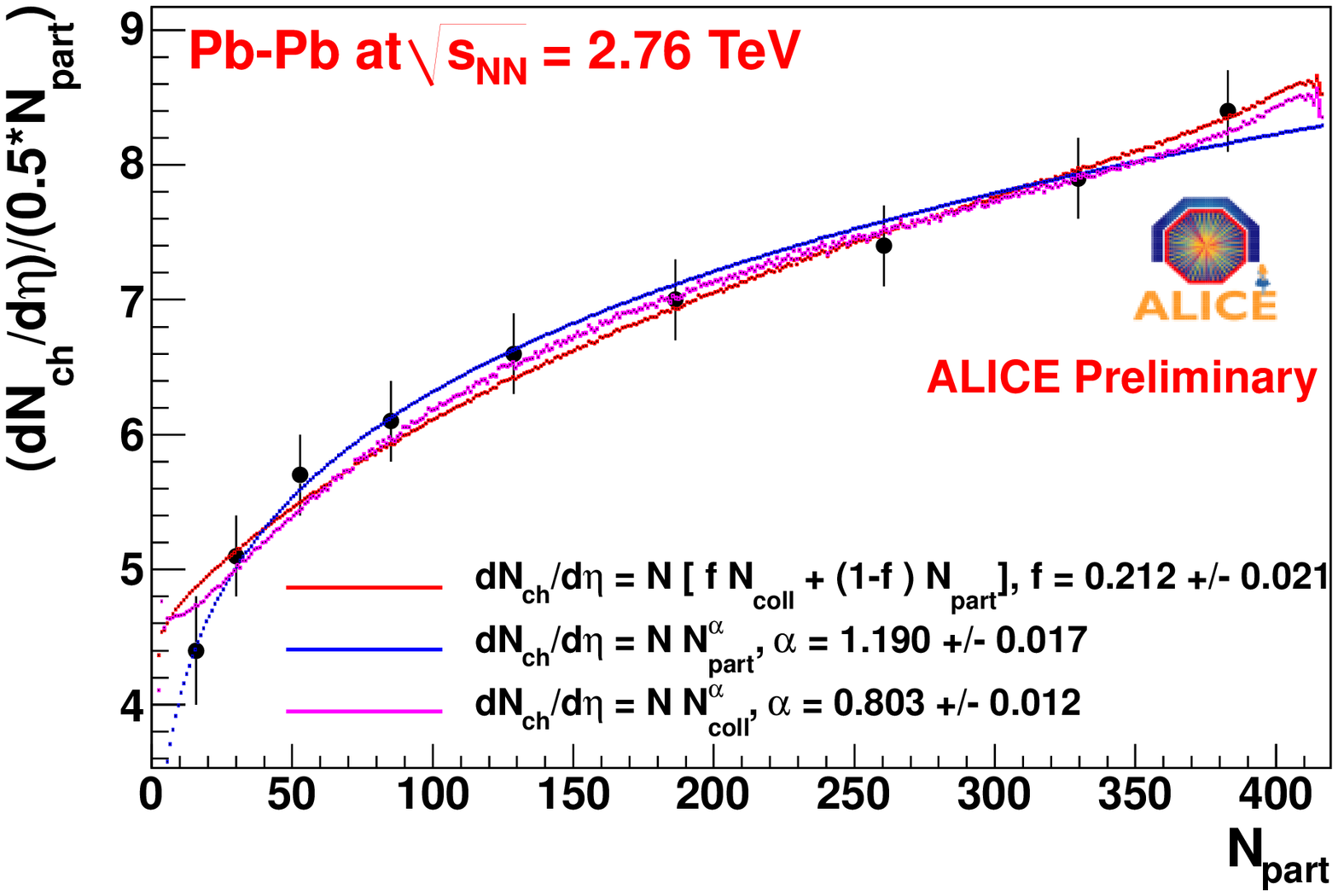}
\caption{
Fits of typical parametrizations for the centrality dependence of $\dNdeta$ per participant
pair to the \PbPb\ data at $\snn=2.76$~TeV. 
}
\label{fig5}
\end{minipage}
\hspace{0.5cm}
\begin{minipage}[t]{0.45\linewidth}
\centering
\includegraphics[width=\textwidth]{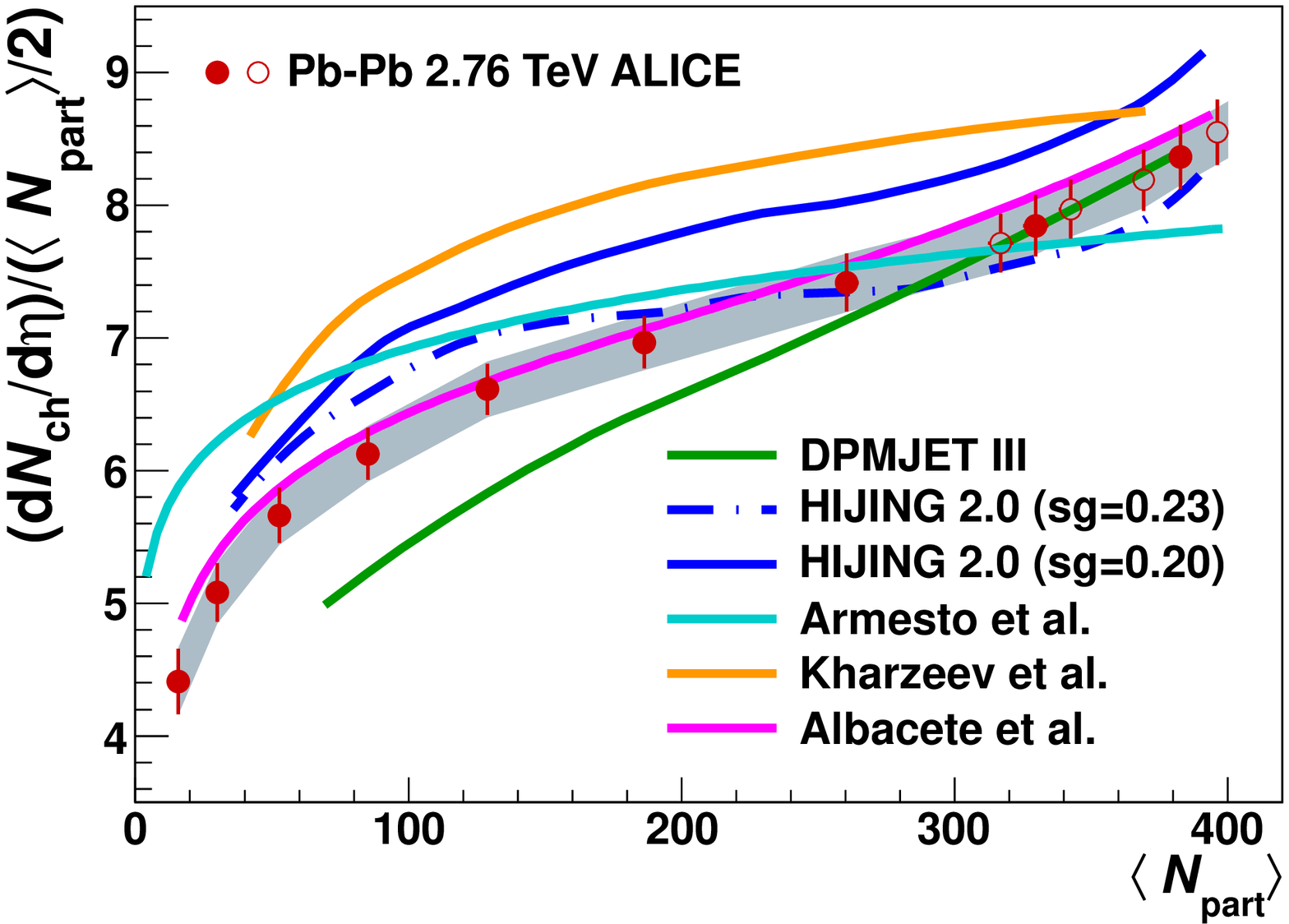}
\caption{
Comparison of $\dNdeta$ per participant pair with model calculations for \PbPb\ at $\snn=2.76$~TeV. 
The figure is from \cite{Aamodt:2010cz}, where the references to all model calculations can be found.}
\label{fig6}
\end{minipage}
\end{figure}

The measurements of $\dNdeta$ and $\dEdeta$ per participant pair as a function of centrality
in \PbPb\ collisions at $\snn=2.76$~TeV show a steady increase by a factor of $2$ between 
peripheral and central collisions~(Figs.\ \ref{fig1} and \ref{fig2}). 
The centrality dependence is similar to that observed in \AuAu\ collisions at $\snn=0.2$~TeV, 
while the yield per participant pair increases by a factor of $2.1$ and $2.5$ for $\dNdeta$ and 
$\dEdeta$, respectively.
The latter is consistent with an increase of the average transverse momentum by about 20\% 
from top RHIC to LHC collision energies.
Taking into account the measurements at lower energy, the energy dependence of both, 
$\dNdeta$ and $\dEdeta$, for the $0$--$5$\% most central collisions phenomenologically 
exhibits a power law scaling, which is stronger than the logarithmic scaling suggested by 
the lower-energy data alone~(Figs.\ \ref{fig3} and \ref{fig4}).
The Bjorken energy density for central collisions, estimated based on $\dEdeta$,
increases by about a factor of $3$ to $\epsilon\tau=16$~GeV/fm$^2$c at $\snn=2.76$~TeV.
Theoretical descriptions of particle production in nuclear collisions are typically either 
based on two-component models combining perturbative QCD processes with soft interactions, 
or saturation models with various parametrizations for the energy and centrality 
dependence of the saturation scale, motivating simple parametrizations of
$\dNdeta$ with $\Npart$ and $\Ncoll$.
The centrality dependence of $\dNdeta$ is well described by three different types of 
such parametrizations with $\Npart$ and $\Ncoll$ revealing that no unique physics conclusion 
can be drawn from such fits~(\Fig{fig5}).
Theoretical descriptions that include a moderation of the multiplicity evolution 
with centrality are favoured by the data~(\Fig{fig6}). 
Since $E_{\rm T}$ also depends on particle composition and their momentum distributions,
the measurement of $\dEdeta$ puts further constraints on models and may lead to
better discrimination.

In summary, the measurements of the centrality dependence of the charged-particle multiplicity density 
and transverse energy at mid-rapidity in \PbPb\ collisions at $\snn=2.76$~TeV have been presented.
Their centrality dependence is found to be strikingly similar to that of the \AuAu\ data at $\snn=0.2$~TeV,
while their yield per participant pair increases by a factor of $2.1$ and $2.5$, respectively.

\vspace{-0.4cm}\enlargethispage{0.5cm}


\section*{References}
\begin{thebibliography}{10}
\bibitem{Aamodt:2010pb}
  K.~Aamodt {\it et al.} [ ALICE Collaboration ],
  Phys.\ Rev.\ Lett.\  {\bf 105}, 252301 (2010).
\bibitem{Aamodt:2010cz}
  K.~Aamodt {\it et al.} [ ALICE Collaboration ],
  Phys.\ Rev.\ Lett.\  {\bf 106}, 032301 (2011).
\bibitem{martin}
  M.~Poghosyan for the ALICE Collaboration, these proceedings.
\bibitem{Adler:2004zn}
  S.~S.~Adler {\it et al.}  [PHENIX Collaboration],
  Phys.\ Rev.\  C {\bf 71}, 034908 (2005)
  [C {\bf 71}, 049901 (2005)]
\bibitem{Adams:2004cb}
  J.~Adams {\it et al.}  [STAR Collaboration],
  Phys.\ Rev.\  C {\bf 70}, 054907 (2004)
\bibitem{Eskola:2001bf}
  K.~J.~Eskola, P.~V.~Ruuskanen, S.~S.~Rasanen, K.~Tuominen,
  Nucl.\ Phys.\  {\bf A696}, 715-728 (2001).
\end{thebibliography}
\end{document}